\documentclass[a4paper,twocolumn,showpacs,pra,nofootinbib]{revtex4-1} 
\usepackage[T1]{fontenc}
\usepackage[latin9]{inputenc}
\usepackage{times}
\usepackage{color}
\usepackage{xspace}
\usepackage{color,ulem}
\usepackage{amsmath}
\usepackage{amssymb}
\usepackage{amsbsy}
\usepackage{graphicx}
\usepackage{bm}
\usepackage{epstopdf}
\usepackage{float}

\usepackage{relsize}

\newcommand{\diff}{\mathrm{d}}

\newcommand{\br}{\ensuremath{\mathbf{r}}\xspace}

\newcommand{\bx}{\ensuremath{\mathbf{x}}\xspace}
\newcommand{\x}{\mathbf{x}}

\renewcommand{\r}{\mathbf{r}}

\newcommand{\bk}{\mathbf{k}}

\begin{document}
\newcommand{\nt}{\tilde{n}}
\newcommand{\Lnh}{\ensuremath{\mathcal{L}}\xspace}
\newcommand{\dbx}{\diff\bx}
\newcommand{\dbk}{\diff\bk}
\newcommand{\nbe}{\ensuremath{\bar{n}_{\mathrm{BE}}}\xspace}

\newcommand{\bro}{\mathbf{r}_1}
\newcommand{\brt}{\mathbf{r}_2}

\newcommand{\ph}{\hat\Psi}
\newcommand{\phd}{\hat\Psi^\dagger}

\newcommand{\pro}{{\phi_L(\rho_1)}}
\newcommand{\prt}{{\phi_L(\rho_2)}}
\newcommand{\dd}{\mathrm{D}}
\newcommand{\deh}{\hat\delta}
\newcommand{\dehd}{\hat\delta^\dagger}

\title{Finite resolution fluctuation measurements of a trapped Bose-Einstein condensate}


\author{R.~N.~Bisset$^{\,1,2}$, C.~Ticknor$^{\,2}$, and P.~B.~Blakie$^{\,1}$}  
\affiliation{1. Jack Dodd Centre for Quantum Technology, Department of Physics, University of Otago, Dunedin, New Zealand\\
2. Center for Nonlinear Studies and Theoretical Division, Los Alamos National Laboratory, Los Alamos, New Mexico 87545, USA}

\begin{abstract}
We consider the fluctuations in atom number that occur within finite-sized measurement cells in a trapped Bose-Einstein condensate (BEC). This approximates the fluctuation measurements made in current experiments with finite resolution \textit{in situ} imaging. A numerical scheme is developed to calculate these fluctuations using the quasiparticle modes of a cylindrically symmetric three-dimensionally trapped condensate with either contact or dipole-dipole interactions (DDIs).
We use this scheme to study the properties of a pancake shaped condensate using cylindrical cells. 
The extension of the theory to washer shaped cells with azimuthal weighting is made and used to discriminate between the low energy roton modes in a dipolar condensate according to their projection of angular momentum. Our results are based on the Bogoliubov approach valid for zero and small finite temperatures. 
\end{abstract}
\pacs{67.85-d,67.85.Bc}

\maketitle


\section{Introduction}
Many important properties of degenerate quantum gases are revealed through their  fluctuations (e.g.~see \cite{Burt1997a,Naraschewski1999,Tolra2004a,Altman2004a,Schellekens2005a,Greiner2005a,Rom2006,Trebbia2006a,Gerbier2006a,Jeltes2007a,Donner2007a,Muller2010a,Hung2011,Guarrera2011a,Hodgman2011a,Jacqmin2011a,Sanner2011a,Armijo2012a,Marzolino2013a,Blumkin2013a,Schley2013a}).
Recently experiments with trapped Bose gases have measured density fluctuations in quasi-one-dimensional (quasi-1D) \cite{Jacqmin2011a,Armijo2012a}, quasi-two-dimensional (quasi-2D)  \cite{Hung2011,Hung2011a} and three-dimensional systems \cite{Blumkin2013a,Schley2013a} using \textit{in situ} absorption imaging.  
Such measurements are necessarily made with finite resolution, so that effectively the number of atoms within a finite-sized cell is counted. The  size of this cell  is typically set by the pixel size of the charge-coupled device used to collect the image and the resolution limited spot size of the imaging system (e.g.~see \cite{Hung2011a}). In practice the smallest spot sizes obtained are comparable to the healing length (typically $\!~\!\sim1\,\mu$m). The size of the cell relative to the physical length scales of the system (e.g.~thermal wavelength and healing length) is crucial in determining the measurement results \cite{Klawunn2011}. 

\begin{figure}[h]
\begin{center}
\includegraphics[width=3.5in]{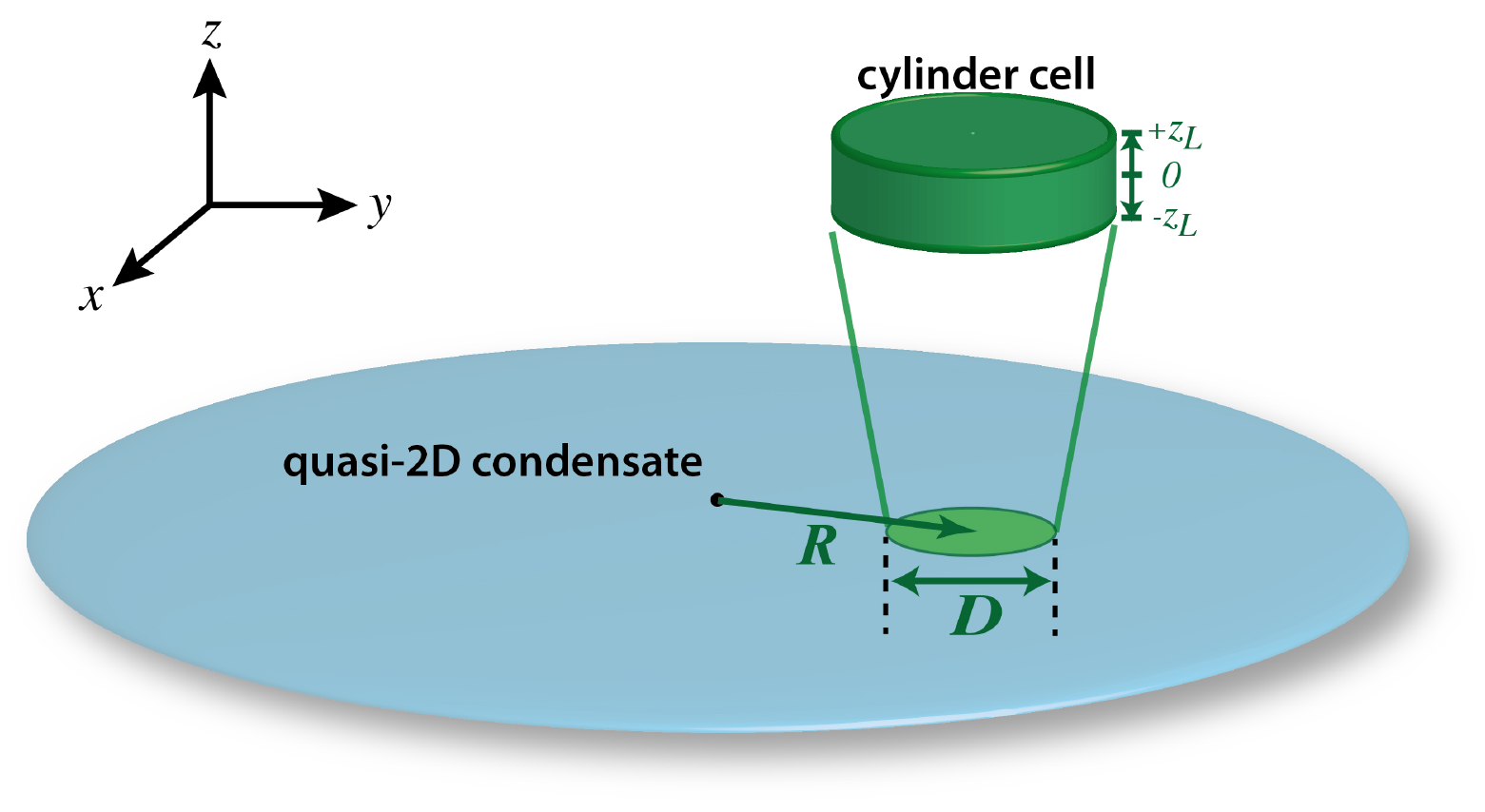}
\caption{(Color online) Schematic   geometry of the system we consider: a radially symmetric pancake condensate realized by tight confinement along $z$; and a cylinder-shaped cell in which  number fluctuations are measured. The cylindrical cell is parameterized by the radius $R$ from trap centre to the cylinder axis, and its diameter $D$.  The $z$ extent (from $-z_L$ to $z_L$) is taken to be larger than the condensate thickness. 
 \label{Fig:Geometry}}
\end{center}
\end{figure}

In this paper we develop a technique to calculate the  fluctuations of a BEC, in which we explicitly include the properties of the finite-sized measurement cells. Recent work \cite{Klawunn2011} has addressed how this can be done in uniform systems. Here we consider the extension   to the trapped system. This situation is rather more challenging because the system lacks translational invariance and the excitations need to be determined numerically.  We apply our scheme to consider the kinds of fluctuation measurements that could be made with \textit{in situ} imaging through a pancake shaped condensate, which should be approximately described by calculating the number fluctuations within cylindrical cells, as schematically shown in Fig.~\ref{Fig:Geometry}. Our results, for condensates with both contact and DDIs, demonstrate that  fluctuation measurements are acutely  sensitive to the nature of the interactions in the system. Finally, we consider  a more complex weighted-washer shaped cell, which can be built up by amalgamating a number of smaller cylindrical cells. We show how this washer cell can be used to reveal the low energy excitations with particular values of angular momentum projection.

The structure of the paper is as follows: 
In section \ref{Sec:Formalism} we outline our formalism for the fully three-dimensional trapped system and detail our numerical scheme for  calculating number fluctuations within  cylindrical cells, as depicted in Fig.~\ref{Fig:Geometry}.
In Sec.~\ref{SEC:Results} we present our main results. We characterize the two-point second order density-density correlation function and then compute the fluctuations within cylindrical cells for a pancake condensate with either contact or DDIs. We then introduce a weighted washer shaped cell and discuss how it might be realized in experiments. We extend our numerical scheme to treat this case and then apply it to characterizing the low energy roton modes of a dipolar BEC. Finally, we
conclude in Sec.~\ref{Sec:Conclusion}.

\section{Formalism}\label{Sec:Formalism}
\subsection{Condensate and quasiparticles}\label{Sec:CondQPs}
Here we consider atoms confined to a cylindrically symmetric harmonically potential 
\begin{equation}
V(\x) = \frac{1}{2}M\omega_\rho^2(\rho^2+\lambda^2z^2),
\end{equation}
where $M$ is the atomic mass and   $\lambda=\omega_z/\omega_\rho$ is the trap aspect ratio. We examine the properties of condensates in which the atoms interact with  DDIs or short ranged contact interactions.  
 For the dipolar  case we take the dipoles to be  polarized along $z$ so that the interaction potential is of the form
\begin{equation} 
U_{\mathrm{D}}(\mathbf{r}) =\frac{3g_{\rm dd}}{4\pi}\frac{1-3\cos^2\theta}{|\mathbf{r}|^3}, \label{Eq:PseudoPot}
\end{equation} 
where   $g_{\rm dd} = \mu_0\mu_m^2/3$, with $\mu_m$ the magnetic dipole moment and $\theta$ is the angle between $\mathbf{r}$ and the $z$ axis. This choice ensures that the system remains cylindrically symmetric, simplifying the numerical treatment.
 For the case of contact interactions $U_{\mathrm{C}}(\mathbf{r})=g\delta(\mathbf{r})$, where  $g=4\pi a_s\hbar^2/M$, with $a_s$ the $s$-wave scattering length. We note that in general dipolar gases interact with both  parts, $U=U_{\mathrm{D}}+U_{\mathrm{C}}$, although the contact part can be tuned using a Feshbach resonance (e.g.~see \cite{Koch2008a}). 

The condensate wavefunction $\psi_0$, which we take to be normalized to the condensate atom number $N_0$, satisfies the nonlocal dipolar Gross-Pitaevskii equation \cite{Goral2000a}
\begin{equation} 
\mu\psi_0 =\left[\!-\frac{\hbar^2\nabla^2}{2M}\!+\!V(\x)\!+\!\int d^3\x^\prime U(\x\!-\!\x^\prime)n_0(\x')\right]\psi_0 ,
\end{equation}
where  $\mu$ is the chemical potential and $n_0(\x')=| \psi_0(\x^\prime)|^2$ is the condensate number density. Our interest is in ground state solutions where the condensate wavefunction can be taken to be real.

 To understand the fluctuations of this system we consider the  field operator, which can be decomposed as
 \begin{equation}
 \hat\Psi(\x) \approx \psi_0(\x) + \hat{\delta}(\x), \label{psihatbog}
\end{equation}
where $\hat{\delta}(\x)$ is the non-condensate operator, with $\langle \hat\delta(\x)\rangle=0$.  The non-condensate operator can be expressed in a quasiparticle expansion as
\begin{equation}
\hat{\delta}(\x)=\sum_i\left[u_i(\x)\hat\alpha_i - v_i^*(\x)\hat\alpha_i^\dagger\right], \label{deltaBog}
\end{equation}
where the amplitude  functions $\{u_i,v_i\}$, with energy eigenvalues $\epsilon_j$, are obtained by solving nonlocal Bogoliubov-de Gennes equations \cite{Ronen2006a}. The quasiparticle operators satisfy $[\hat{\alpha}_i,\hat{\alpha}_j^\dagger]=\delta_{ij}$, and in thermal equilibrium have an occupation given by
\begin{equation}
\bar{n}_j\equiv \langle\hat{\alpha}_j^\dagger\hat{\alpha}_j\rangle=\frac{1}{e^{\beta\epsilon_j}-1},
\end{equation}
where $\beta=1/k_BT$ is the inverse temperature. The non-condensate number density is given by
\begin{align}
\tilde{n}(\x)&=\langle\delta^\dagger(\x)\hat{\delta}(\x)\rangle,\\ 
&=\sum_j\left[|u_j(\x)|^2\bar{n}_j+|v_j(\x)|^2(\bar{n}_j+1)\right],\label{ntilde_den}
\end{align}
so that the total density is
\begin{equation}
n(\x)=n_0(\x)+\tilde{n}(\x).
\end{equation}

We solve for the condensate and quasiparticles exploiting the cylindrical symmetry of the problem. For example, the quasiparticles can be calculated in sub-spaces of fixed $z$-projection of angular momentum ($m_j$) as $u_j(\x)=u_j(\rho,z)e^{im_j\phi}$, where $\rho$ and $\phi$ are the radial and angular coordinates for the $x$-$y$ plane. The numerical solutions are then found on special quadrature grids based on Bessel ($\rho$ direction) and Fourier ($z$ direction) decompositions, which differ for each value of angular momentum  projection \cite{Ronen2006a}. Further details on how we solve the Gross-Pitaevskii and Bogoliubov-de Gennes equations can be found in Ref.~\cite{Blakie2013}

\subsection{Number fluctuations within cells}
We initially focus on the case of cylindrically shaped measurement cells (see Fig.~\ref{Fig:Geometry}) where we denote the region inside the cell as $\sigma$.
The number fluctuations within the cell may be characterized by the variance of the atom number,
\begin{equation}
\delta N^2_\sigma \equiv  \Big\langle \big( \hat{N}_\sigma- N_\sigma \big)^2 \Big\rangle , \label{Eq:NumFluc}
\end{equation}
where
\begin{equation}
\hat{N}_\sigma \equiv \int_\sigma d^3\x \,\phd(\x)\ph(\x),
\end{equation}
 is the cell number operator, and $N_\sigma\equiv\langle \hat{N}_\sigma \rangle$. 
The mean number of atoms within the measurement cell $\sigma$ is given by
\begin{equation}
N_{\sigma}=N_{0\sigma}+\tilde{N}_\sigma,\label{Nsigma}
\end{equation}
where
\begin{align}
N_{0\sigma}&=\int_\sigma d^3\x\,n_0(\x),\\
\tilde{N}_{\sigma}&=\int_\sigma d^3\x\,\tilde{n}(\x),
\end{align}
are the mean condensate and non-condensate number in the cell, respectively.

The fluctuations in number about the mean is  
 \begin{align}
 \delta N_\sigma^2 &=  \int_\sigma d^3\x_1\!\int_\sigma d^3\x_2 \,S_{nn}(\x_1,\x_2),\label{dN2}
\end{align}
where $S_{nn}(\x_1,\x_2)\equiv \langle  \delta \hat{n}(\x_1)\delta\hat{n}(\x_2)\rangle$ is the Ursell density-density correlation function, and we have introduced the density fluctuation operator
\begin{align}
\delta \hat{n}(\x) &\equiv \hat{\Psi}^\dagger(\x) \hat{\Psi}(\x)-n(\x),\\
&\approx\psi_0(\x)\left[\hat{\delta}(\x)+\hat{\delta}^\dagger(\x)\right].\label{dn_bog}
\end{align}
In the last line we have used Eq.~(\ref{psihatbog}) and neglected the small terms which are second order in the quasiparticles. This should be a good approximation for temperatures well below the condensation temperature and away from the condensate surface. Both of these conditions ensure that $n_0\gg\tilde{n}$. Evaluating Eq.~(\ref{dN2}) using (\ref{dn_bog}) gives
 \begin{align}
 \delta N_\sigma^2 &\!=  \!\sum_i\!\int_\sigma d^3\!\x_1\!\int_\sigma d^3\x_2 \,\delta n_i(\x_1)\delta n_i^*(\x_2)\coth\left(\frac{\beta\epsilon_i}{2}\right),\label{dN2b}
\end{align} 
where $\delta n_i(\x) \equiv \psi_0(\x)\left[u_i(\x) -v_i(\x) \right]$ is the density fluctuation amplitude of the $i$-th quasiparticle. 

\subsubsection{Uniform system}\label{Sec:Uniform}
In this section we consider a homogeneous condensate  in $\dd$ spatial dimensions. We take the system to be in a box of  volume $V=L^\dd$, and subject to periodic boundary conditions, so that the condensate is $\psi_0=\sqrt{n_0}$,   with $n_0=N_0/L^\dd$  the condensate density. 
In this case the cell fluctuations are \cite{Yukalov2005a,Astrakharchi2007a,Klawunn2011}
 \begin{align}
 \delta N_\sigma^2 &= n \int \frac{d^\dd\mathbf{k}}{(2\pi)^\dd}\,\tilde\tau(\mathbf{k})S(\mathbf{k}),\label{dN2_homog}
\end{align}
where $n=N/L^{\mathrm{D}}$ is the total density and 
\begin{equation}
S(\mathbf{k}) = \frac{1}{N}\sum_j\left|\int d^\dd\x\,e^{-i\mathbf{k}\cdot{\x}}\psi_0(\x)[u_j(\x)-v_j(\x)]\right|^2,
\end{equation}
is the static structure factor. The function $\tilde\tau(\mathbf{k})$  describes the cell geometry, and is  the Fourier transform of $\tau(\br)\equiv \int_\sigma d^\dd\x_1\int_\sigma d^\dd\x_2\,\delta(\br+\x_2-\x_1)$ (see \cite{Klawunn2011} for details). 

We note that for the uniform system the  Ursell correlation function is translationally invariant, i.e. $S_{nn}(\x_1,\x_2)\to S_{nn}(\x_1-\x_2)$,  and in obtaining Eq.~(\ref{dN2_homog}) we have used
\begin{equation}
S(\mathbf{k}) = \frac{1}{n}\int d^\dd\br\, S_{nn}(\br)e^{-i\bk\cdot\br}.
\end{equation} 
For a uniform condensate  the structure factor is given by the Feynman relation
\begin{equation}
S(\mathbf{k})=\alpha_f\frac{\epsilon_0(\bk)}{\epsilon(\bk)}\coth\left(\frac{\beta\epsilon(\bk)}{2}\right),\label{SkBog}
\end{equation}
where $\epsilon_0(\bk)=\hbar^2k^2/2M$, $\alpha_f=n_0/n$, 
\begin{equation}
\epsilon(\bk)=\sqrt{\epsilon_0(\bk)[\epsilon_0(\bk)+2n_0\tilde{U}(\bk)]},\label{BogDisp}
\end{equation}
is the Bogoliubov dispersion relation, and $\tilde{U}(\bk)$ the Fourier transform of the interaction potential $U(\r)$. We note that   Eq.~(\ref{SkBog}) is valid for small condensate depletion, i.e.~$\alpha_f\approx1$.

While result (\ref{dN2_homog}) does not apply to the trapped case, it does provide a starting point for a local-density treatment of trapped systems, which we develop further in Sec.~\ref{SEC:LDA}.

\subsubsection{Trapped system}
For the trapped system, using modes obtained by numerical diagonalization, it is inconvenient to calculate the Ursell function because it has a shot-noise part that behaves  as $\delta(\x_1-\x_2)$, and  thus slowly converges as we increase the number of modes. Instead it is convenient to normally order the operators to obtain 
\begin{equation}
\delta N^2_\sigma = N_\sigma + \int_\sigma\int_\sigma  G^{(2)}(\x_1,\x_2) d^3\x_1 d^3\x_2 - N_\sigma^2 , \label{Eq:FlucG2}
\end{equation}
where
\begin{equation}
G^{(2)}(\x_1,\x_2) \equiv \langle\phd(\x_1) \phd(\x_2)\ph(\x_1) \ph(\x_2)\rangle, \label{Eq:G2}
\end{equation}
is the normally ordered density-density correlation function.

Making use of Wick's theorem \cite{BlaizotRipka,Holzmann1999} and keeping only terms to second order in the quasiparticle operators we arrive at \footnote{Note that this approximation is valid when $n_0(\x) \gg \tilde{n}(\x), |\tilde{m}(\x)|$.}
\begin{widetext}
\begin{align}
\delta N_\sigma^2&=  N_\sigma + \,\int_\sigma\int_\sigma d^3\x_1 d^3\x_2\, \psi_0(\x_1)\psi_0(\x_2)\left[ \tilde{m}(\x_1,\x_2) + \tilde{n}(\x_1,\x_2) + \mathrm{c.c}\right] , \label{Eq:NFluc}        
\end{align}
where 
\begin{align}
\tilde{n}&(\x_1,\x_2) = \left\langle \dehd(\x_1)\deh(\x_2) \right\rangle \label{Eq:ndep} =  \sum_j \left[ u_j^*(\x_1)u_j(\x_2)\bar{n}_j + v_j(\x_1)v_j^*(\x_2)(1+\bar{n}_j) \right], \\
\tilde{m}&(\x_1,\x_2) = \left\langle \deh(\x_1)\deh(\x_2) \right\rangle = -\sum_j \left[ u_j(\x_1)v_j^*(\x_2)(1+\bar{n}_j) + v_j^*(\x_1)u_j(\x_2)\bar{n}_j \right],
\end{align}
are the non-condensate and  anomalous density matrices, respectively \footnote{Both $\tilde{n}$ and $\tilde{m}$ are real since the sum is taken over all modes.} (also see \cite{Yukalov2009a}). 
The local character of these quantities [i.e.~$\tilde{n}(\x)=\tilde{n}(\x,\x)$ and $\tilde{m}(\x)\equiv\tilde{m}(\x,\x)$] was investigated for  systems with contact and DDIs in Ref.~\cite{Blakie2013}.
We emphasize that Eq.~(\ref{Eq:NFluc}) is equivalent to Eq.~(\ref{dN2b}), but converges much faster with the number of quasiparticle modes included in the summation.
\end{widetext}

\subsubsection{Local Density Approximation}\label{SEC:LDA}
The uniform system treatment of Sec.~\ref{Sec:Uniform} can be extended to the trapped system using a local density approximation (LDA). The LDA theory has also been used to obtain the dynamic structure factor for  trapped BECs (e.g.~see \cite{Zambelli2000,Blakie2002a}), relevant to Bragg spectroscopy. Because our interest in this paper is in pancake shaped systems we formulate this section making a quasi-2D approximation, i.e.~utilizing the results in Sec.~\ref{Sec:Uniform} with $\dd=2$. To do this we integrate out the $z$-dimension to obtain the 2D density profile of the  condensate $n_{2D}(\bm{\rho})\equiv\int dz\,|\psi_0(\bm{\rho},z)|^2$, where $\bm\rho=(x,y)$ is the in-plane position vector. An important issue is the form of the Fourier transformed interaction $\tilde{U}(\bk_{\bm{\rho}})$ for the quasi-2D system. The approach we use is discussed in Refs.~\cite{Blakie2012a,Bisset2013} and we do not repeat here.

The essence of the LDA approach is to treat each part of the trapped system as a locally homogeneous system at the same density, and then add up all such contributions in the region of interest (i.e.~cell). This approach should be valid where the uniform system treatment presented in Sec.~\ref{Sec:Uniform} is valid ($n_0\gg\tilde{n})$ and where the density of the system does not vary rapidly with position. 

 We compute the averaged static structure factor for the system within the cell $\sigma$  by the density weighted LDA sum 
\begin{equation}
S_{\sigma}(\mathbf{k}_{\rho}) = \frac{1}{N_\sigma}\int_{\sigma} d^2\bm{\rho}\,n_{2D}(\bm{\rho})  S(\mathbf{k}_{\rho},\bm{\rho}),
\end{equation}
 where $ S(\mathbf{k}_{\rho},\bm{\rho})$ is the uniform static structure factor (\ref{SkBog}) evaluated using the density at $\bm\rho$ [i.e.~$ n_0\to n_{2D}(\bm{\rho})$]\footnote{Note we also take $\alpha_f=1$ in applying the uniform treatment, and only make use of the condensate density.}. The number fluctuations for the trapped system is then given by Eq.~(\ref{dN2_homog}) with the replacement $S(\mathbf{k}_{\rho})\to S_{\sigma}(\mathbf{k}_{\rho})$. Additionally, we note that under the quasi-2D approximation the cylindrical cells become disks in the $\bm\rho$-plane, and the geometry function appearing in Eq.~(\ref{dN2_homog}) is \cite{Klawunn2011} 
 \begin{equation}
 \tilde{\tau}(\bk_{\bm{\rho}})=\pi^2D^2J_1^2(\tfrac{1}{2}Dk_\rho)/k_{\rho}^2,
 \end{equation}
 where $J_1$ is the Bessel function, and $D$ is the disk diameter.

Details of our full algorithm for evaluating the fluctuations according to Eq.~(\ref{Eq:NFluc}) are given in the Appendix.

\section{Results}\label{SEC:Results}

\subsection{Systems and parameters}\label{Sec:sysmandparams}
\begin{figure} 
\begin{center} 
\includegraphics[width=3.3in]{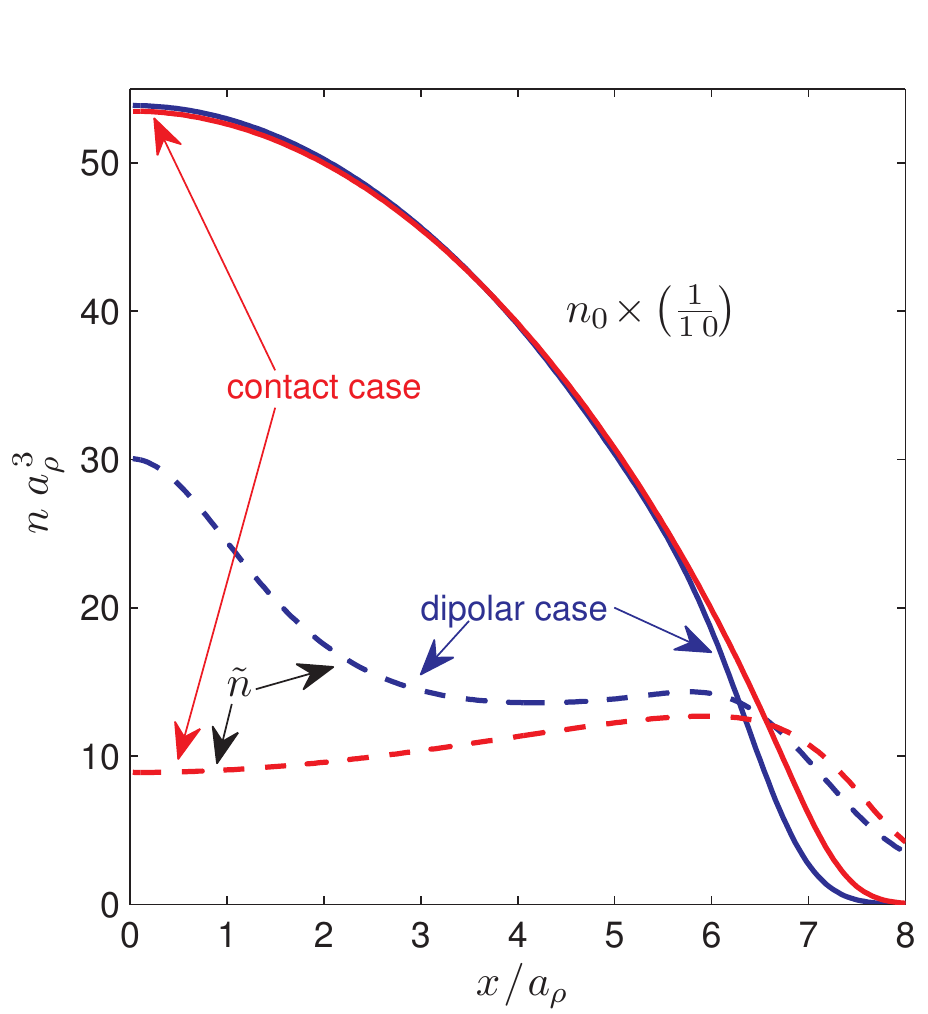} 
\caption{(Color online)  Condensate (solid line) and non-condensate (dashed line) density for a system with purely DDIs (blue/black)  and purely contact interactions (red/grey). The condensate density is scaled by a factor of $\tfrac{1}{10}$ to allow easier comparison to the non-condensate density. Other parameters $T = 10\,\hbar\omega_\rho/k_B$, $N_0 = 25\times10^3$ and $\lambda=20$. For the purely dipolar case $D_{\mathrm{I}} = 220$, while for the solely contact case $C_{\mathrm{I}}  = 127$.
 \label{Fig:dencomp}}
\end{center}
\end{figure}

For our results we define length in terms of the radial harmonic oscillator length $a_\rho = \sqrt{\hbar/M\omega_\rho}$, and following Ref.~\cite{Ronen2006a} we utilize the dimensionless interaction parameters $C_{\mathrm{I}} = N_0a_s/a_\rho$ and $D_{\mathrm{I}} = 3N_0g_{\mathrm{dd}}M/4\pi\hbar^2a_\rho$  for the contact and DDIs, respectively.

We calculate results for  $N_0=25\times 10^3$ atoms within a pancake shaped trap with aspect ratio $\lambda = 20$. 
In the results we focus on the comparison of  two parameter sets: (i) a system interacting purely via a contact interaction, i.e.~with $C_{\mathrm{I}}  = 127$ and $D_{\mathrm{I}}  = 0$; (ii) a system interacting purely via a DDI, i.e.~with  $C_{\mathrm{I}}  = 0$ and $D_{\mathrm{I}} = 220$. Both parameter sets have been chosen so that the condensates have nearly the same chemical potential $\mu = 37.6\,\hbar\omega_\rho$ \footnote{In the pancake geometry the DDI is predominantly repulsive and the effect on the chemical potential is overall positive.}. This ensures that many properties of the two systems are comparable. For example, in Fig.~\ref{Fig:dencomp} we compare the density profiles of the two systems along the $x$-axis, demonstrating that the density and shapes of the condensates are very similar.
To put the dipolar parameters into the context of current experiments the choice of $^{164}$Dy, with dipole moment $\mu_m=10\,\mu_B$, would correspond to a radial trapping frequency $\omega_\rho = 2\pi\times 10.8~$s$^{-1}$.

For the choice of parameters we make here the dipolar condensate has roton like excitations \cite{Santos2003a}. These roton modes arise in regimes of tight confinement along the direction that the dipoles are polarized, and for sufficiently strong DDIs. In this case  the effective interaction in the $x$-$y$ plane  becomes momentum dependent \cite{Fischer2006a}. For in-plane momenta $\hbar k_{\rho}\lesssim\hbar/a_z$, where $a_z=\sqrt{\hbar/M\omega_z}$ is the $z$-confinement length,  the DDI is repulsive, while for $\hbar k_{\rho}\gtrsim\hbar/a_z$ it crosses over to being attractive. Thus modes of wavelength  $\lambda_{\mathrm{rot}}\!\sim\!a_z$ can be energetically softened by their interaction with the condensate, in which case we refer to them as rotons. Interestingly these modes are sensitive to the condensate density and, due to the radial trapping, they are effectively confined as a \textit{roton gas}  in the high density central part of the condensate \cite{JonaLasinio2013}. Detailed properties of the rotons for the parameter regime we consider here can be found in Refs.~\cite{Blakie2013,Bisset2013b}.

\subsection{Two Point Correlations}\label{Sec:2pCorr}
\begin{figure} 
\begin{center}
\includegraphics[width=3.3in]{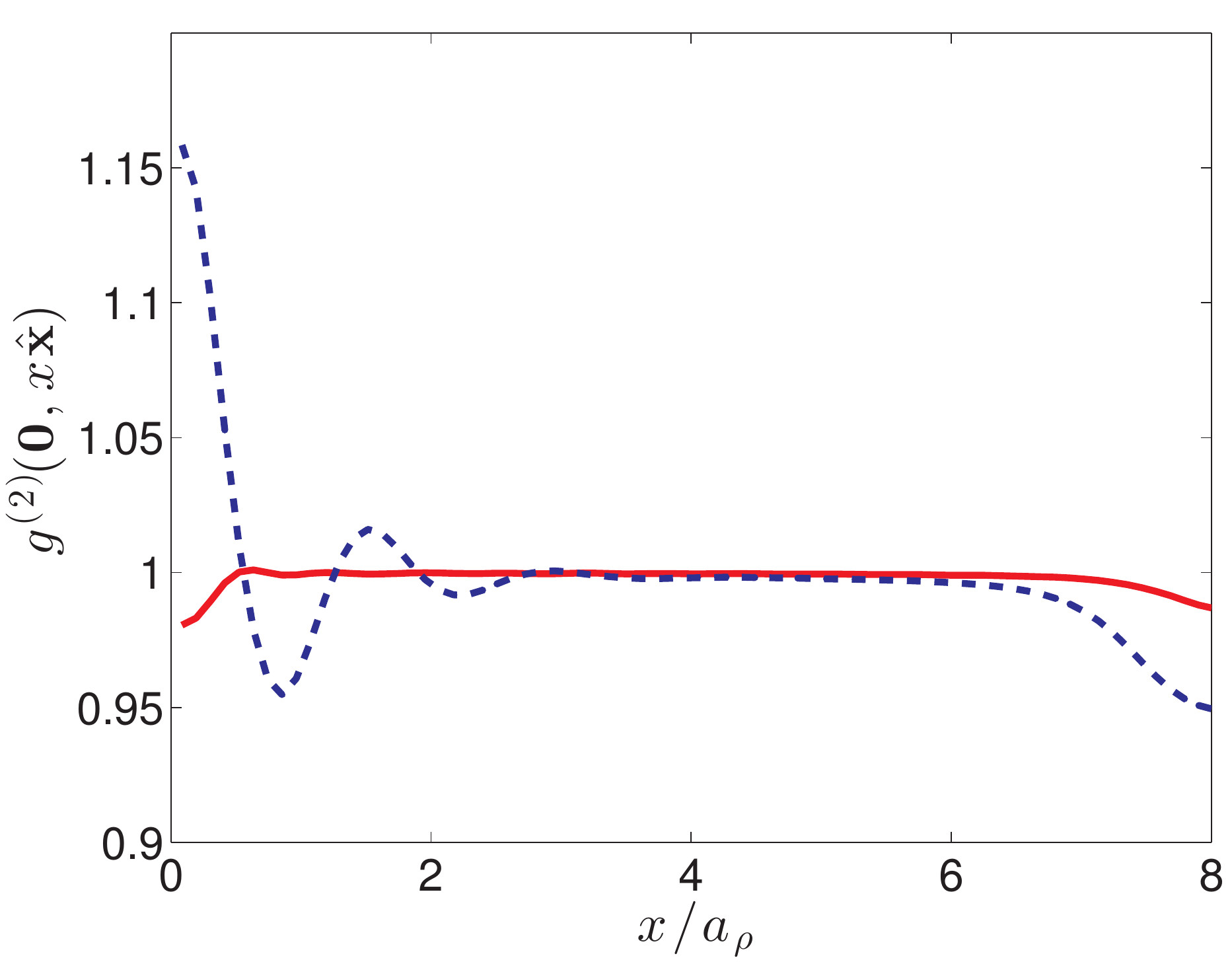} 
\caption{(Color online) The normalized correlation function $g^{(2)}({\bf 0},x\hat{\bf{x}})$ for a condensate with purely dipole interactions (dashed line) and purely contact interactions (solid line). 
  Other parameters $T = 10\,\hbar\omega_\rho/k_B$, $N_0 = 25\times10^3$ and $\lambda=20$. For the purely dipolar case $D_{\mathrm{I}} = 220$, while for the solely contact case $C_{\mathrm{I}}  = 127$.
 \label{Fig:G2}}
\end{center}
\end{figure}

We first examine the two-point normally ordered density correlation function, with results comparing the purely dipolar and contact condensates presented in Fig.~\ref{Fig:G2}. Here we show the normalized correlation function \cite{Naraschewski1999} defined by
\begin{equation}
g^{(2)}(\mathbf{x}_1,\mathbf{x}_2) = \frac{G^{(2)}(\mathbf{x}_1,\mathbf{x}_2)}{n(\mathbf{x}_1)n(\mathbf{x}_2)},
 \end{equation} 
 with one point taken at trap center $\x_1=\mathbf{0}$ and the other taken  to lie along the $x$-axis, $\x_2=x\hat{\x}$. Previous work has considered aspects of this correlation function for condensates with contact interactions \cite{Lee1957,Dodd1997,Holzmann1999,Blakie2005a,Bezett2008a,Wright2011a}, and DDIs \cite{Ticknor2012,Sykes2012} in the quasi-two-dimensional regime.

For the case of pure contact interactions    $g^{(2)}(\mathbf{0},x\hat{\x})$ is near unity everywhere that the condensate density is significant, except for a dip occurring for small point separation ($x\lesssim0.5\,a_{\rho}$).
The dip is a manifestation of suppressed density fluctuations due to repulsive contact interactions and occurs on a length scale set by the healing length $\xi=\hbar/\sqrt{Mgn}$, which is $\xi\approx 0.17 \,a_\rho$ for our system (using peak density at trap centre).

In contrast, the case of pure dipolar interactions demonstrates a marked peak for small separation and subsequent oscillations with increasing $x$. The large central peak is due to the attractive character of the DDI for the short wavelength roton modes. The choice of the fixed point at the origin results in oscillations dominated by the $m_j=0$ roton modes as these are the only excitations with non-zero density at the trap center \footnote{Note that the fixed point is actually at the center-most radial grid point which is not precisely at the origin, this implies a small contribution from $|m_j|>0$ modes.}.  As noted in Sec.~\ref{Sec:sysmandparams}, the roton wavelength is approximately set by the $z$ confinement length, and is  $\lambda_{\mathrm{rot}}\approx 1.5\,a_{\rho}$ [from Fig.~\ref{Fig:FlucCorrCells}(b)].
The reduction in $g^{(2)}$ for $x\gtrsim7\,a_\rho$ occurs because at these positions the non-condensate density begins to exceed the condensate density (see Fig.~\ref{Fig:dencomp}) which invalidates our evaluation of (\ref{Eq:G2}).

\subsection{Cylinder cell fluctuations}\label{Sec:ResultsNumFluc}
 \begin{figure} 
\begin{center}
\includegraphics[width=3in]{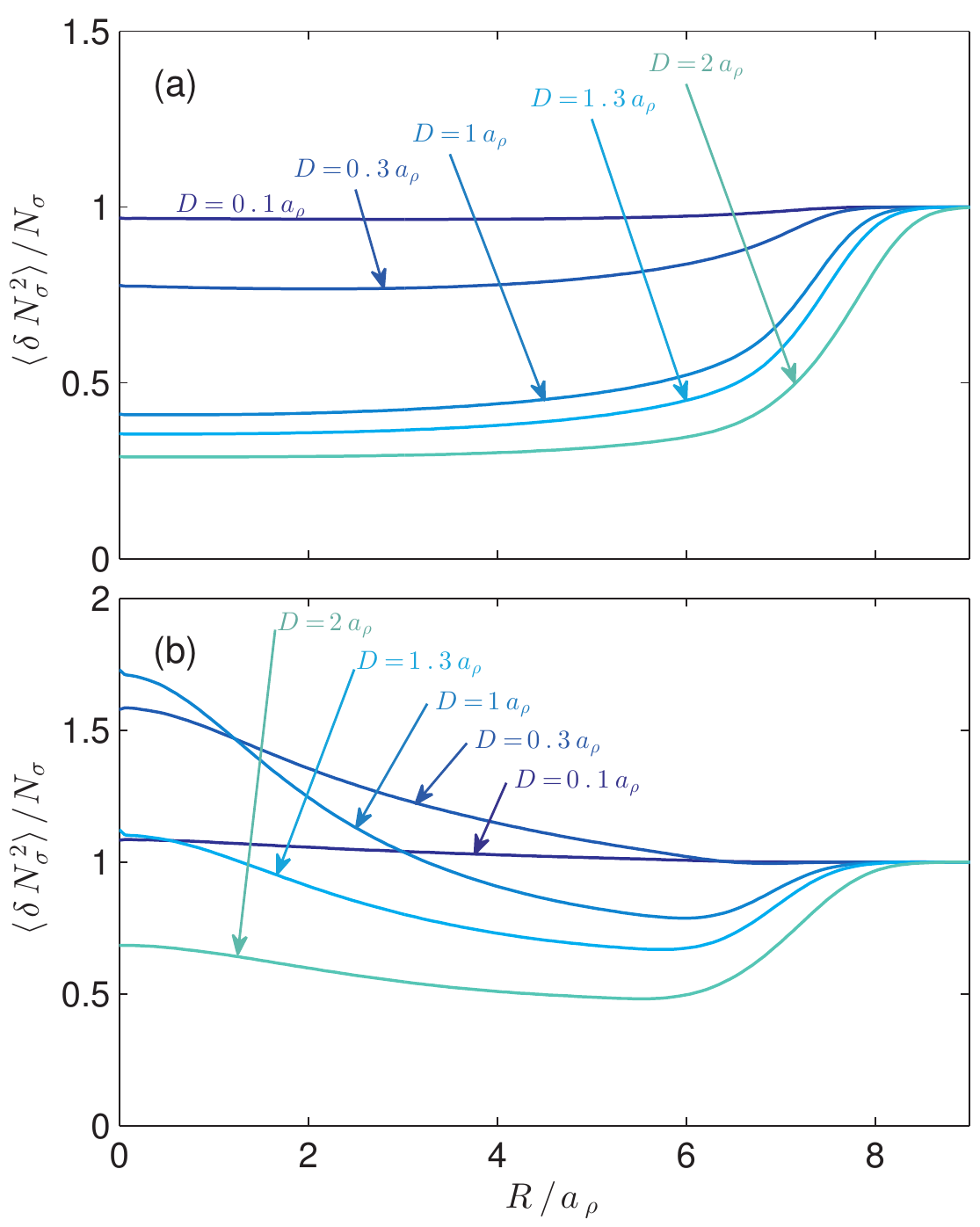}
\caption{(Color online) Number fluctuations as a function of cylinder position $R$ for various diameters $D$ at zero temperature for (a) pure contact interactions   $C_{\mathrm{I}}$ = 127 and (b) pure dipolar interactions $D_{\mathrm{I}}$ = 220. Other parameters as in Fig.~\ref{Fig:G2}.
 \label{Fig:FlucCellsT0}}
\end{center}
\end{figure}

 \begin{figure} 
\begin{center}
\includegraphics[width=3in]{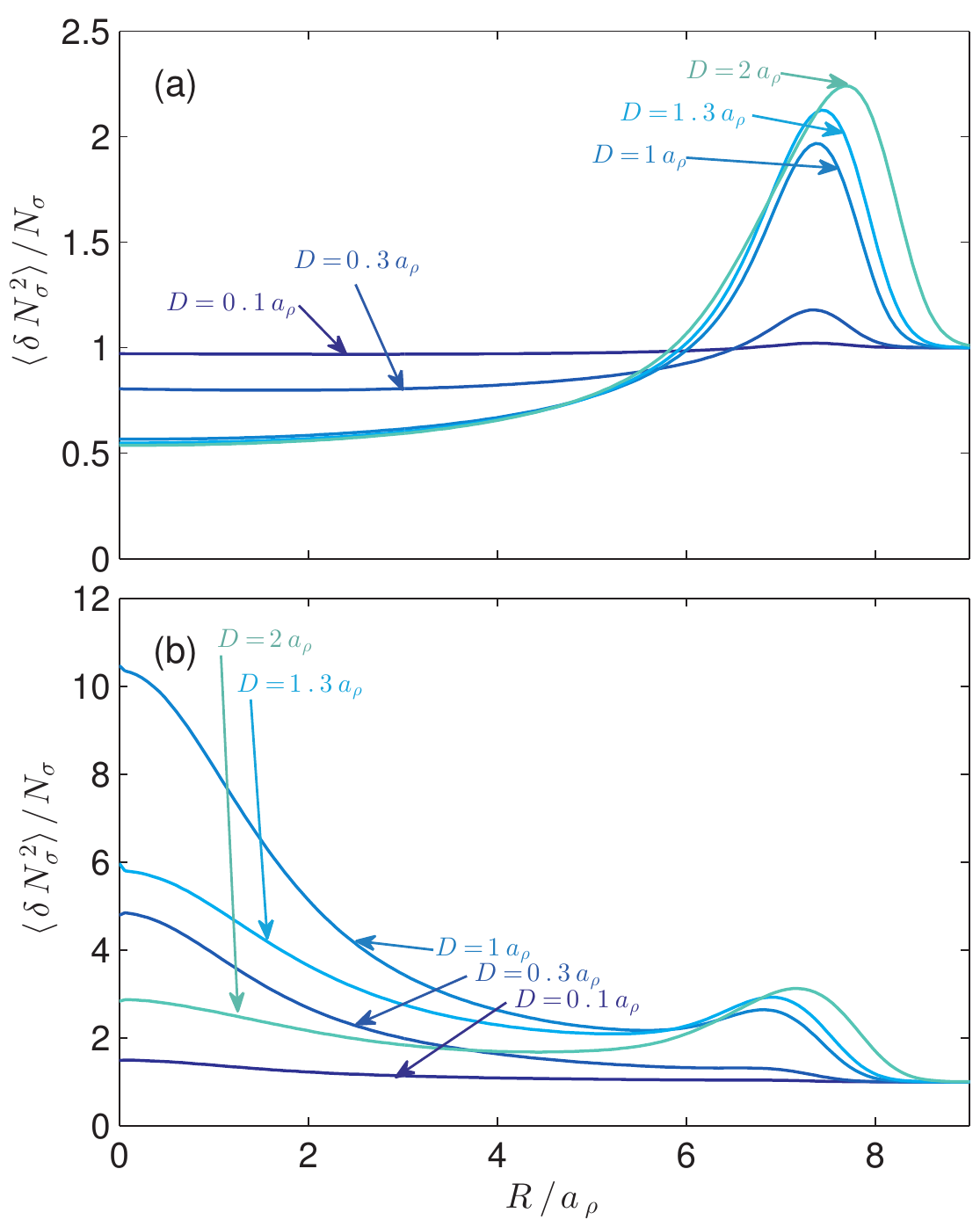}
\caption{(Color online) Number fluctuations as a function of cylinder position $R$ for various diameters $D$ at $T=10\,\hbar\omega_\rho/k_B$ for (a) pure contact interactions $C_{\mathrm{I}}$ = 127 and (b) pure dipolar interactions $D_{\mathrm{I}}$ = 220. Other parameters as in Fig.~\ref{Fig:G2}.
 \label{Fig:FlucCellsT10}}
\end{center}
\end{figure}

\subsubsection{zero temperature}
In Fig.~\ref{Fig:FlucCellsT0} we investigate the role of interactions on number fluctuations within cylindrical cells at zero temperature. 
The purely contact case is examined in Fig.~\ref{Fig:FlucCellsT0}(a). These results reveal that the fluctuations are sub-poissonian ($\delta N_\sigma^2<N_\sigma$) in the central region where the condensate is dense, and that the relative fluctuations ($\delta N_\sigma^2/N_\sigma$) decrease with increasing cell size. For large cells  ( $D\gg\xi$)  the fluctuations are sensitive to the long wavelength (phonon) modes of the system qualitatively consistent with the treatment developed in Ref.~\cite{Klawunn2011} for the uniform  quasi-2D Bose gas.
For cells near the surface of the condensate ($R\sim8\,a_\rho$), where the density is low and interaction effects reduce, the fluctuations approach poissonian.

For the pure dipolar case the relative fluctuations exhibit a non-monotonic dependence on cell size.  Notably, near the trap center  a super-poissonian peak emerges for intermediate cylinder sizes that best match the roton modes, i.e.~$D=1\,a_\rho$ for $R\approx0$ \cite{Bisset2013,Klawunn2011}, where this value of $D$ is roughly half the  roton wavelength.  
As these cells are are moved outwards (e.g. $D=1\,a_\rho$ for $R\gtrsim3\,a_\rho)$ the relative fluctuations decrease.  This reveals the narrow confinement of the rotons to the high density region of the condensate \cite{JonaLasinio2013}.

\subsubsection{non-zero temperature} 
We assess the influence of temperature in Fig.~\ref{Fig:FlucCellsT10}, by considering the same systems presented in Fig.~\ref{Fig:FlucCellsT0}, but with the temperature increased to  $T = 10\,\hbar\omega_\rho/k_B$ (corresponds to $5.3$ nK, which is about 14\% of the condensation temperature $T_c$). For pure contact interactions [Fig.~\ref{Fig:FlucCellsT10}(a)] two significant changes occur: (i)  The fluctuations increase near the surface $R\approx 6a_\rho$, where the non-condensate density is largest
and the suppression effect of the repulsive interactions is weakest (also noting that for $\rho>7\,a_\rho$ is also where $\tilde{n}\gtrsim n_0$   and our approach is invalid).
(ii) The fluctuations in large cells ($D\gtrsim1\,a_\rho$) tend to increase. This occurs because the large cells are dominated by phonon modes which are thermally activated at the temperature considered. In contrast  the smaller cells ($D<1\,a_\rho$) are dominated by shorter wavelength excitations that are largely frozen out, and the fluctuations remain near their $T=0$ value. The behavior we observe in Fig.~\ref{Fig:FlucCellsT10}(a) is qualitatively similar to the behavior seen in Fig.~1 of Ref.~\cite{Armijo2012a}, where density fluctuations were measured for a quasi-1D Bose gas.

Similarly, in Fig.~\ref{Fig:FlucCellsT10}(b) we observe that temperature tends to increase the fluctuations of the purely dipolar condensate [c.f.~Fig.~\ref{Fig:FlucCellsT0}(b)].
Notably, a large roton \textit{peak} occurs at trap centre for cells with $D\sim1\, a_{\rho}$. This enhancement is due to the thermal activation of the rotons, and  is considerably more prominent than the peak in two point correlations function observed for the same parameters in  Fig.~\ref{Fig:G2}.

\subsubsection{Comparison to LDA approach}
 \begin{figure} 
\begin{center} 
\includegraphics[width=3in]{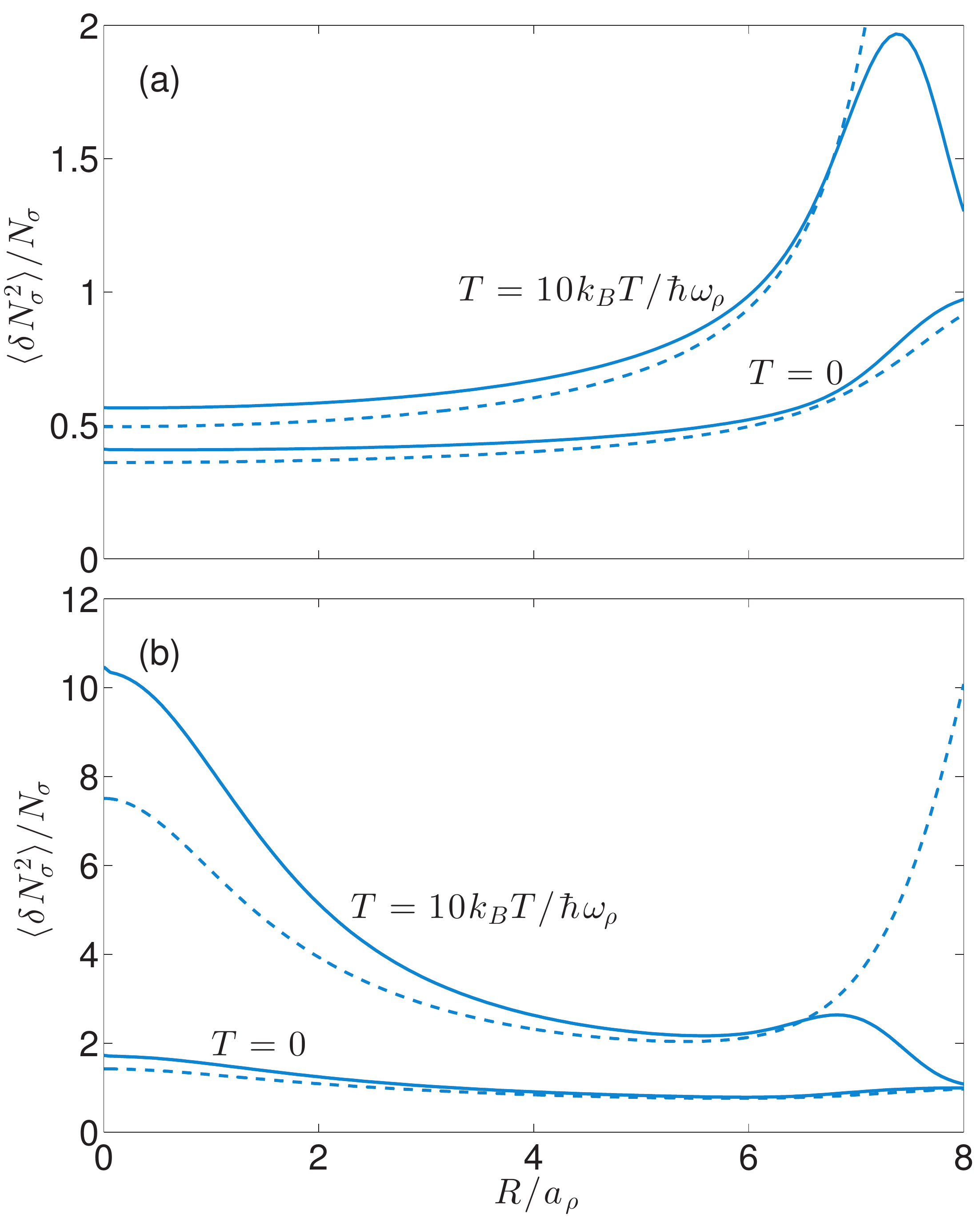} 
\caption{(Color online) Cylindrical cell fluctuations from full 3D (solid) and LDA (dashed) calculations for both $T=0$ and $T=10\, \hbar\omega_\rho/k_B$ with a cell diameter of $D = 1a_\rho$. (a) Pure contact case   $C_{\mathrm{I}}= 127$. (b) Pure dipolar case $D_{\mathrm{I}}=220$.
 \label{Fig:LDA}}
\end{center}
\end{figure}
 
In Fig.~\ref{Fig:LDA} we compare the full Bogoliubov calculations against the LDA approach outlined in Sec.~\ref{SEC:LDA}. Because the LDA approach only requires the condensate mode $\psi_0$ [to obtain $n_{2D}(\bm{\rho})$ and $\tilde{U}(\bk_\rho)$] it is much simpler to implement, and avoids the complexity of solving for all of the quasiparticle modes.

The most noticeable difference is at $R\gtrsim 7\,a_{\rho}$ where the LDA result starts to increase rapidly with $R$. This occurs at the surface of the condensate where both approaches are invalid.  We also note that for the dipolar case [Fig.~\ref{Fig:LDA}(b)] the agreement for $R\lesssim3\,a_{\rho}$ is less satisfactory at finite temperatures  than what we observe for the contact case [Fig.~\ref{Fig:LDA}(a)].  This arises because the roton modes, which are thermally activated at this temperature, are not well described within the LDA. Temperature in the dipolar case also tends to increase the non-condensate density at trap center (relative to the contact case, see Fig.~\ref{Fig:dencomp}).

\subsection{Weighted-washer shaped cell fluctuations}\label{Sec:WeightedCells}

 \begin{figure}[h]
\begin{center}
\includegraphics[width=3.2in]{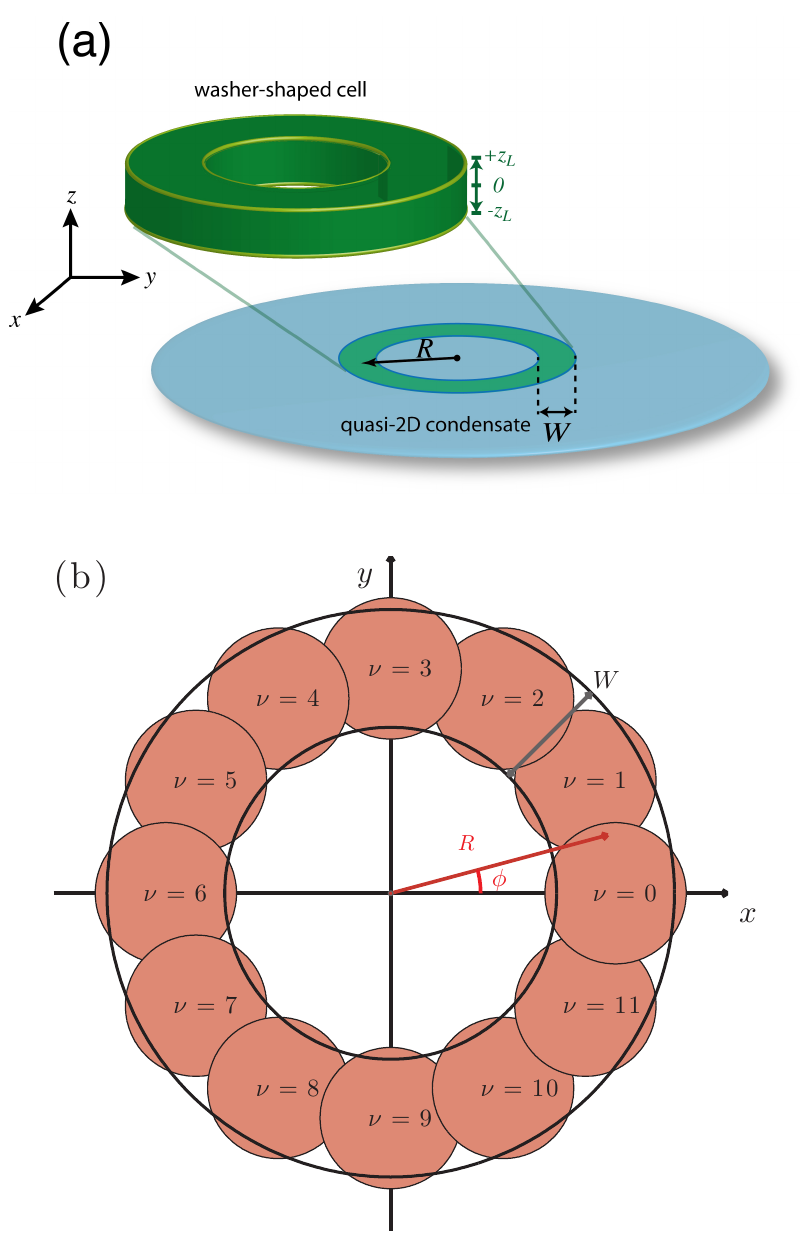}
\caption{(Color online) (a) The geometry of the washer-shaped cell.  This cell is centered on the trap center, and is parameterized by the mid-radius   $R$ and  width $W$, with a $z$ extent  taken to be larger than the condensate thickness. 
(b)  In practice the washer cell can be constructed using  a set of cylinder sub-cells, which we have labeled by $\nu$.
A weighted cell measurement can be made by multiplying the number measurement made in each sub-cell by a phase factor $e^{-im\phi_\nu}$, where $\phi_\nu$ is the angle to the center of  sub-cell $\nu$.
 \label{Fig:wcells}}
\end{center}
\end{figure}
\subsubsection{Measurement operator}
We define the washer shaped cell $\sigma$ to be concentric with the trap center, and to have a mid-radius $R$ and radial thickness $W$ [see Fig.~\ref{Fig:wcells}(a)].  We require $R> W/2$, to ensure that the washer has a finite central core. Like the cylinder cell we assume that the $z$ extent is larger than the condensate thickness. 

In Ref.~\cite{Bisset2013} it was found that considering fluctuations within washer-shaped cells centered at the origin selects for the contribution from $m_j = 0$ modes. Here we generalize this idea by defining a weighted number operator for washer-shaped cell $\sigma$ as
\begin{equation}
\hat{N}_{\sigma m} \equiv \int_\sigma d^3 \x\, e^{-im\phi} \phd(\x)\ph(\x),
\end{equation}
with $m$ being an integer.  We note that effectively  $\hat{N}_{\sigma m}$ is the azimuthal Fourier transform of the density operator in the region $\sigma$.
In practice the washer-shaped cell could be formed by amalgamating measurements from a number of smaller cylindrical sub-cells, weighted by appropriate phases, as indicated in Fig.~\ref{Fig:wcells}(b). 
Recently experiments have demonstrated similar types of combined cell correlation measurements:   In Ref.~\cite{Hung2013} a Fourier transform over the cells was used to correlate density fluctuations at different locations in a  quasi-2D condensate; In Ref.~\cite{Armijo2012a} cells were amalgamated in the analysis of a quasi-1D condensate to verify the scaling of fluctuations with cell size.

The fluctuations in $\hat{N}_{\sigma m}$ are given by
\begin{equation}
\delta N^2_{\sigma m} \equiv \left\langle  |\hat{N}_{\sigma m}-{N}_{\sigma m}  |^2\right\rangle, \label{Eq:FlucWeighted}
\end{equation}
where ${N}_{\sigma m}  =\langle\hat{N}_{\sigma m} \rangle$ is the mean value, with  ${N}_{\sigma m}=0$ for $m\ne0$, due to the cylindrical symmetry of the system.
The numerical algorithm for evaluating this is a minor variation to the cylinder-shaped cell algorithm and details  are given in the Appendix.

\subsubsection{Application to dipolar condensate in roton regime}
 \begin{figure}
\begin{center}
\includegraphics[width=3.2in]{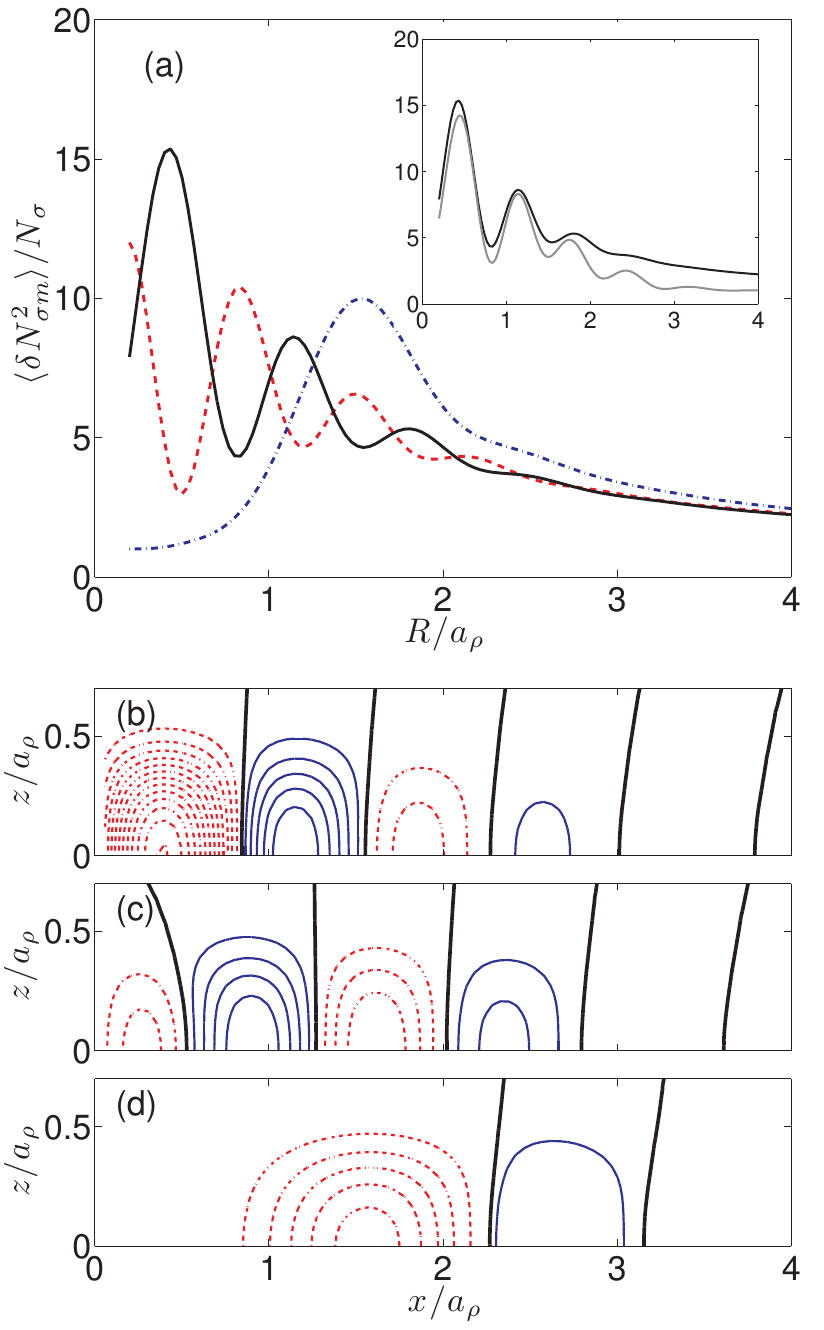}
\caption{(Color online) Weighted number fluctuations within washer-shaped cells for $D_{\mathrm{I}}$ = 220, $C_{\mathrm{I}}$ = 0 and $T = 10\,\hbar\omega_\rho/k_B$ with washer width $W = 0.4a_\rho$. (a) $m$ = 0 (dashed), 1 (solid) and 6 (dot-dashed). Inset: Reproduction of $m$ = 1 result (black) along with the contribution solely due to the two $m_j = 1$ roton modes (gray). (b),(c) Contours of the density fluctuation amplitude $\delta n_i$ [see Eq.~(\ref{dN2b})] for the two roton modes considered in the inset for subplot \ref{Fig:FlucCorrCells} (a). These modes have the third (b) and sixth (c) to lowest energies within the $m_j = 1$ subspace. (d) The lowest roton mode for $m_j=6$, having the second to lowest energy in the $m_j=6$ subspace. Solid lines are positive and dashed lines are negative with thick black lines indicating the zero crossings; contour spacing  $3/a_\rho^3$. Other parameters as in Fig.~\ref{Fig:G2}.
 \label{Fig:FlucCorrCells}}
\end{center}
\end{figure}

In Fig.~\ref{Fig:FlucCorrCells}(a) we evaluate the weighted-washer cell fluctuations for a dipolar condensate, presenting results for  $m = 0, 1$ and $6$. 
We note that the \textit{unweighted} case $m=0$ was considered previously in Ref.~\cite{Bisset2013}. There it was noted that the fluctuations in the washer cells varied rapidly with cell  radius $R$ as compared to the case of a cylindrical cell [c.f.~Fig.~\ref{Fig:FlucCellsT10}(b)]. This observation  also holds true for the $m=1$ case, but the variation is largely reduced for the $m=6$ case. This variation arises because the washer cell selects the contributions from a few quasiparticles of particular angular momentum projection. If these modes are rotonic modes they contribute strongly to the fluctuation signal \cite{Blakie2013} and tend to have a rapid spatial variation.  
To confirm that the  fluctuation measurements are dominated by a few roton modes  we compare the full ($m = 1$) calculation with that using only the two lowest energy $m_j=1$ roton excitations [i.e.~the sum in Eq.~(\ref{dN2m}) restricted to these two modes]  \footnote{The rotons for the parameters of the calculation we present here are extensively characterized in Ref.~\cite{Bisset2013b}, particularly Figs.~1 and 3(b).} and show this comparison in the inset. The agreement is particularly good near the trap center where these modes are confined, as can be seen by the roton profiles shown in Figs.~\ref{Fig:FlucCorrCells}(b) and (c). In contrast for the $m=6$  case the relevant roton mode propagates azimuthally and only has a weak radial dependence [Figs.~\ref{Fig:FlucCorrCells}(d)] and the flucutation signal in in Fig.~\ref{Fig:FlucCorrCells}(a) varies much more slowly as $R$ changes. Also the larger angular momentum of this mode forces it further from the trap center.

\section{Conclusions and outlook}\label{Sec:Conclusion}
In this paper we have developed a method for calculating the number fluctuations within finite-sized cells of a trapped BEC. We have investigated the canonical case of cylindrical cells in a pancake shaped BEC, which is relevant to current experiments using absorption imaging. We present results for condensates with purely contact or purely dipole interactions. These results reveal the strong dependence of the fluctuations on the nature of the interactions. We also use these results to validate a simpler LDA approach which is found to have reasonable agreement with the full solutions.

We propose the idea of using  weighted-washer shaped cells, which could be formed from a set of underlying cylindrical sub-cells. The fluctuations measured in the washer cell discriminates contributions from particular classes of quasiparticles modes, i.e.~those with a particular angular momentum projection, with the particular projection values selected being determined by the choice of weighting. We demonstrate that this scheme can probe roton modes within a dipolar condensate, and that the measured fluctuation signal comes from the lowest few modes. Rotons have not yet been identified in experiments and fluctuation measurements complement a number of other proposals for  schemes to detect these modes (e.g.~see \cite{Corson2013a,*Corson2013b,*JonaLasinio2013b}).
The importance of discriminating  rotons with  $m_j\ne0$ has been identified by Ronen \textit{et al.}~\cite{Ronen2007a}: In certain parameter regimes they predict that the dipolar condensate will spontaneously take a bi-concave (density oscillating) shape, and that accompanying this transition  the lowest energy roton will change from having angular momentum projection $m_j=0$ (``radial roton") to $m_j\ne0$ (``angular roton").  More recently it was shown that angular rotons can be engineered by applying an external potential to force the condensate density into a bi-concave shape \cite{Bisset2013b}. We note however, that when the trap is not cylindrically symmetric $m_j$ is no longer a good quantum number and the roton modes distort \cite{Martin2012a}.  

The high resolution imaging required to experimentally measure the fluctuations  in small cells is a capability that is now appearing in many experiments. At the forefront of such work Hung \textit{et al.}~\cite{Hung2013} have developed a system to calculate the static structure by such \textit{in situ} measurements of density fluctuations. In those experiments the maximum wavevector for the structure factor was limited to $\approx 2\,\mu m^{-1}$, due to their resolution limit. This setup would immediately carry over to the construction of weighted-washer shape cells, with  the maximum value of $m$  limited to the washer circumference divided by the imaging system resolution limit. 

Natural extensions of this work are to go beyond the Bogoliubov approach in calculating the quasiparticle modes. That is to include the   back-action of the non-condensate atoms on themselves and the condensate e.g., using Hartree-Fock or other similar approximations \cite{Ronen2007b,Bisset2012,Ticknor2012a}. In the case of dipolar gases, where the fluctuations due to the roton modes become significant, these corrections may be important even at low temperatures \cite{Boudjemaa2013a}.
 
\section*{ Acknowledgments}  We acknowledge fruitful discussions with J.~Armijo. P.~B.~B. and R.~N.~B. acknowledge support by the Marsden Fund of New Zealand (contract UOO1220).
R.~N.~B. and C.~T. acknowledge support from CNLS, LDRD, and LANL which is operated by LANS, LLC for the NNSA of the US DOE (contract no.~DE-AC52-06NA25396).

\appendix
\section*{Appendix: Numerical algorithm}
In this appendix we briefly outline the algorithm we use to accurately and efficiently compute the fluctuations [i.e. evaluate Eq.~(\ref{Eq:NFluc})].
\subsection{Cylinder-shaped cells}\label{Sec:Num}
 \begin{figure*}[tbh!]
\begin{center}
\includegraphics[width=5.7in]{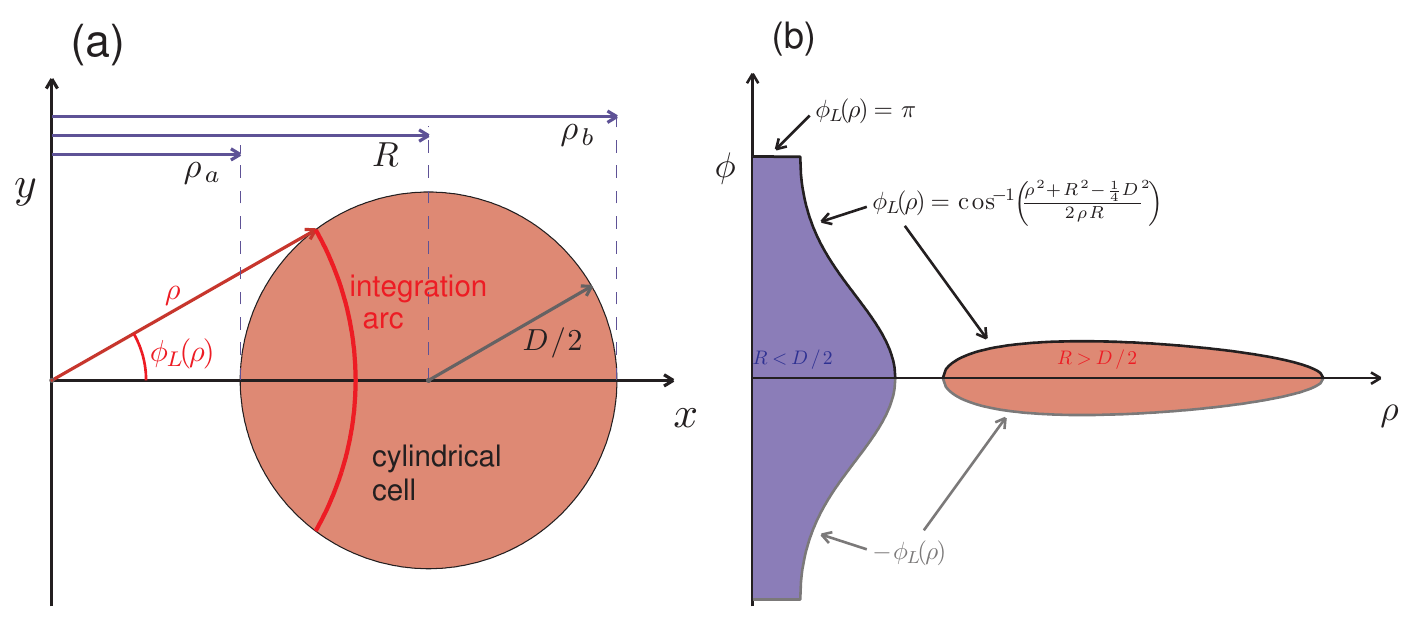}
\caption{(Color online) Schematic showing how we integrate over the cylindrical cell in the $x$-$y$ plane by summing over arcs that vary as a function of $\rho$. (a) A cylinder-shaped cell (shaded) in the $x$-$y$ plane. An integration arc demonstrates the angular limit function $\phi_L(\rho)$. The radial limits $\rho_a$ and $\rho_b$ are also indicated.  
(b) A cylinder-shaped cell (shaded) in the $\rho$-$\phi$ plane. One case with $R>\frac{1}{2}D$ (right shaded region) corresponds to a cylinder cell that does not contain the origin: For this case $\rho_a=R-\frac{1}{2}D$, $\rho_b=R+\frac{1}{2}D$, and $\phi_L(\rho)=\cos^{-1}([\rho^2+R^2-\tfrac{1}{4}D^2]/2\rho R)$ for $\rho_a\le \rho\le\rho_b$. The other case with $R<\frac{1}{2}D$ (left shaded region) is a cylinder cell that contains the origin: For this case $\rho_a=0$, $\rho_b=R+\frac{1}{2}D$, and $\phi_L(\rho)=\pi$ for $\rho\le \tfrac{1}{2}D-R$ and $\phi_L(\rho)=\cos^{-1}([\rho^2+R^2-\tfrac{1}{4}D^2]/2\rho R)$ for $\tfrac{1}{2}D-R\le \rho\le\rho_b$.
\label{Fig:DiskInt}}
\end{center}
\end{figure*}
Taking advantage of the cylindrical symmetry of the problem we  decompose the modes as
\begin{align}
u_j(\x_1) &= e^{i(m_j\phi_1 + S_j)}u_j(\rho_1,z_2),\\
v_j^*(\x_2) &= e^{-i(m_j\phi_2 + S_j)}v_j(\rho_2,z_2),
\end{align}
where  $m_j$ is the  angular momentum projection along $z$ of mode $j$ and $S_j$ represents a constant phase for each mode that   cancels from the observables we compute.
By construction we take two-dimensional functions $u_j(\rho,z)$ and $v_j(\rho,z)$ to be real. 

The integration region $\sigma$ for the cylinder-shaped cell is shown in Fig.~\ref{Fig:Geometry}.   The limits in the $z$-direction ($-z_L$ and $z_L$) are taken to be symmetric about zero and large enough so that the cylinder height is greater than the condensate thickness in this direction. Thus we can focus on the shape of the cell in the $x$-$y$ plane and take all atoms, irrespective of their $z$ coordinate, to contribute (as would be the case of a column density taken with absorption imaging along $z$).
Without loss of generality we position the cell symmetrically about the $x$-axis  so that the limits of $\phi$ integration are symmetric about zero (which we refer to as $\phi_L$ and $-\phi_L$), as shown in Fig.~\ref{Fig:DiskInt}(a) for the case of a cylinder-shaped cell in the $x$-$y$ plane, and in the $\rho$-$\phi$ plane in Fig.~\ref{Fig:DiskInt}(b). 
The $\phi$ integration is performed first and in general $\phi_L$ is $\rho$ dependent, meaning that the $x$-$y$ integration of cylindrical cells is performed by summing arcs that subtend various angles. See Fig.~\ref{Fig:DiskInt} for additional information on the definition of cylinder-shaped cells. 

\begin{widetext}
Returning to our expression for number fluctuations, Eq.~(\ref{Eq:NFluc}) now reads
\begin{align}
&\delta N^2_\sigma = N_\sigma +\sum_j\int_{\rho_a}^{\rho_b} d\rho_1\,\rho_1\int_{-z_L}^{z_L} dz_1\int_{-\pro}^\pro d\phi_1\int_{\rho_a}^{\rho_b} d\rho_2\,\rho_2\,\int_{-z_L}^{z_L} dz_2\int_{-\prt}^\prt  d\phi_2 \,\psi_0(\rho_1,z_1)\psi_0(\rho_2,z_2) \notag \\ 
&\times2\cos(m_j(\phi_1-\phi_2))\Big[ -u_j(\rho_1,z_1)v_j(\rho_2,z_2)\{ 1 + 2\bar{n}_j\}  
+ u_j(\rho_1,z_1) u_j(\rho_2,z_2) \bar{n}_j  
 + v_j(\rho_1,z_1)v_j(\rho_2,z_2)\{1+\bar{n}_j\} \Big] .  \label{Eq:FlucCylin} 
\end{align}
This expression can be simplified considerably to a form suitable for numerical evaluation.
Crucially the $\phi$ integral is analytic, i.e.
\begin{align}
\int_{-\prt}^\prt &\int_{-\pro}^\pro \cos(m_j(\phi_1-\phi_2))d\phi_1 d\phi_2 = \frac{4}{m^2_j}\sin(m_j\pro)\sin(m_j\prt).
\end{align}
\end{widetext}

The mean cell number [see Eq.~(\ref{Nsigma})] is evaluated as
\begin{align}
N_{0\sigma} &= 2\int_{\rho_a}^{\rho_b} \phi_L(\rho) \rho \left[\int_{-z_L}^{z_L} n_0(\rho,z) dz\right] d\rho ,\label{N0s}\\
\tilde{N}_\sigma &=\sum_j\left\{\mathcal{A}_j\bar{n}_j+\mathcal{B}_j\left(1+\bar{n}_j\right)\right\},
\end{align}
where
\begin{align}
\mathcal{A}_j &= 2\int_{\rho_a}^{\rho_b} \phi_L(\rho) \rho  \left[\int_{-z_L}^{z_L} u_j(\rho,z)^2 dz \right] d\rho, \\
 \mathcal{B}_j &= 2\int_{\rho_a}^{\rho_b} \phi_L(\rho) \rho  \left[\int_{-z_L}^{z_L} v_j(\rho,z)^2 dz \right] d\rho.
\end{align}

The number fluctuations can be evaluated as
\begin{align}
\delta N_\sigma^2 =  N_{\sigma} + \sum_j\left\{\mathcal{C}_j (1\!+\!2\bar{n}_j )\!+\!\mathcal{D}_j \bar{n}_j\!+\!\mathcal{E}_j (1\!+\!\bar{n}_j)\right\}, \label{Eq:dN2_CompFriend}
\end{align} 
where
\begin{equation}
\mathcal{C}_j = -8 \alpha_j\beta_j,~~~~\mathcal{D}_j = 8 \alpha_j^2~~~~\mathrm{and}~~~~\mathcal{E}_j = 8 \beta_j^2, \label{Eq:CjDjEj}
\end{equation}
and
\begin{equation}
\alpha_j = \int_{\rho_a}^{\rho_b}\!\rho  \frac{\sin(m_j\phi_L(\rho))}{m_j}\left[\int_{-z_L}^{z_L} u_j(\rho,z)\psi_0(\rho,z) dz\right]d\rho ,\label{alphaj}
\end{equation}
\begin{equation}
\beta_j =  \int_{\rho_a}^{\rho_b}\!\rho  \frac{\sin(m_j\phi_L(\rho))}{m_j}\left[\int_{-z_L}^{z_L} v_j(\rho,z)\psi_0(\rho,z) dz\right]d\rho ,\label{betaj}
\end{equation}
with $ \tfrac{1}{m_j} \sin(m_j\phi_L(\rho))\to\phi_L(\rho)$ for $m_j=0$. 
 
The key quantities to be evaluated, given in Eqs.~(\ref{N0s})-(\ref{betaj}), benefit because the integrations over the $z$-variable can be carried out in advance (these terms are all in square brackets).
We also note that because $u_j$ and $v_j$ are obtained on quadrature grids (see Sec.~\ref{Sec:CondQPs}), careful interpolation onto a specialized \textit{cell-integration} grid is required to avoid boundary effects when calculating the various integrals over the variable $\rho$. This is particularly important when the cell size is small compared to the resolution of the grids on which the condensate and quasiparticles are calculated.
 
In a typical case, which involves using  $\sim10^4$ pre-calculated quasiparticle modes, it takes around 2 minutes per cell to calculate $\delta N^2_\sigma$ on a single core machine. For each of our figures we normally need to consider several hundred different cells which would take of order a day on a single core. However, the calculation for each cell is independent and this process may be efficiently implemented in parallel.

\begin{widetext}
\subsection{Weighted-washer shaped cells}
Our numerical procedure follows that outlined above with a few small modifications. 
The $x$-$y$ plane integration is easily converted to the $\rho$-$\phi$ plane (c.f.~Fig.~\ref{Fig:DiskInt}), where it is given by  $\phi_L=\pi$ for $\rho_a\le\rho\le\rho_b$, with $\rho_a=R-\tfrac{1}{2}W$ and $\rho_a=R+\tfrac{1}{2}W$. 
The fluctuations are then calculated according to 
\begin{align}
\delta N^2_{\sigma m}&= N_{\sigma} +  \sum_j \left[\mathcal{C}_j( 1 + 2\bar{n}_j) +\mathcal{D}_j\,\bar{n}_j + \mathcal{E}_j(1 + \bar{n}_j)\right]\delta_{m,m_j} ,\label{dN2m}
\end{align}
with the replacement $\frac{1}{m_j}\sin(m_j\phi_L)\to\pi$ in Eqs.~(\ref{alphaj}) and (\ref{betaj}). 
The term $\delta_{m,m_j} $ in Eq.~(\ref{dN2m}) shows that a cell symmetrically placed (with respect to the symmetry axis of the system) selects contributions from quasiparticles of a definite angular momentum projection.
\end{widetext}

%
%

%

\end{document}